# Differentiating Photoexcited Carrier and Phonon Dynamics in the Δ, L, and Γ Valleys of Si(100) with Transient Extreme Ultraviolet Spectroscopy

Scott K. Cushing[1,2], Angela Lee[1], Ilana J. Porter[1,2], Lucas M. Carneiro[1,2], Hung-Tzu Chang[1], Michael Zürch[1], Stephen R. Leone[1,2,3,*]

[1]Department of Chemistry, University of California, Berkeley, CA 94720, USA.

[2]Chemical Sciences Division, Lawrence Berkeley National Laboratory, Berkeley, CA 94720, USA.

[3]Department of Physics, University of California, Berkeley, CA 94720, USA.

*Corresponding Author e-mail address: srl@berkeley.edu

**ABSTRACT:** Transient extreme ultraviolet (XUV) spectroscopy probes core level transitions to unoccupied valence and conduction band states. Uncertainty remains to what degree the core-hole created by the XUV transition modifies the measurement of photoexcited electron and hole energies. Here, the Si $L_{2,3}$ edge is measured after photoexcitation of electrons to the Δ, L, and Γ valleys of Si(100). The measured changes in the XUV transition probability do not energetically agree with the increasing electron photoexcitation energy. The data therefore experimentally confirm that, for the Si $L_{2,3}$ edge, the time-dependent electron and hole energies are partially obscured by the core-hole perturbation. A model based on many-body approximations and the Bethe-Salpeter equation is successfully used to predict the core-hole's modification of the final transition density of states in terms of both electronic and structural dynamics. The resulting fit time constants match the excited state electron thermalization time and the inter-valley electron-phonon, intra-valley electron-phonon, and phonon-phonon scattering times previously measured in silicon. The outlined approach is a more comprehensive framework for interpreting transient XUV absorption spectra in photoexcited semiconductors.

## I. INTRODUCTION

Transient extreme ultraviolet (XUV) spectroscopy offers advantages in terms of temporal resolution and element specificity when compared with visible-light and infrared table top sources[1–3]. Unlike traditional UV and IR transient absorption measurements, the XUV probe pulses created by high-harmonic generation can have attosecond pulse widths[4–6]. The XUV probe's energy range allows core level to valence transitions to be measured, with the photoexcited change in absorption containing both electronic and structural dynamics[7]. The combination of temporal resolution and probed energy range possible with XUV sources has led to new insights into the first few-femtoseconds of photoexcitation, as well as into the longer timescale dynamics of electron-phonon and phonon-phonon scattering[8–14]. The same is true for x-ray transient absorption studies performed using x-ray free electron lasers and synchrotrons[15–20].

Interpreting photoexcited dynamics from the differential x-ray absorption is challenging because of the core-hole, which is the positive charge left in the core-level by the core-to-valence XUV probe transition. While the localized core-hole wavefunction is beneficial because it imparts structural information onto the x-ray absorption spectrum, in doing so the core-hole potential also modifies the final transition density of states[21–23]. The change in the final transition density of states masks the energy of photoexcited electrons and holes, sometimes obscuring direct measurement of their time-dependent energies. As a result, the ground state x-ray absorption spectrum and any photoexcited changes seldom match the expected effects from visible and infrared measurements of the same material[24–26]. In the case of localized valence orbitals, such as in the 3d transition metals, the core-hole interaction is strong enough that only the photoexcited change in elemental oxidation state can be measured[27–30].

Recent reports suggest that the core-hole in covalent semiconductors with delocalized valence electrons, like Ge and Si, is negligible enough that the electron and hole energies can be measured as a function of time over a broad XUV energy range[31–34]. It remains less clear to what degree transient XUV spectra can be used to distinguish the occupation of non-degenerate valleys in semiconductor band structures. It also remains unknown whether the unique electron-phonon scatterings pathways that thermalize photoexcited carriers between and within each valley can be quantified from the photoexcited changes in XUV or x-ray absorption.

In this paper, the differential absorption of the Si $L_{2,3}$ edge is measured following visible optical excitation to the $\Delta$, L, and $\Gamma$ valleys of silicon. The changes in the XUV absorption at the critical points do not agree with the expected energies of photoexcited carriers because of core-hole effects. A previously reported model[33] based on the Bethe-Salpeter equation (BSE) with density functional theory (DFT) and simplified by several many-body approximations is tested for analyzing the electron energies within the core-hole perturbation. By fitting the BSE-DFT model to the experimental data, the electron energy, optical phonon population, and acoustic phonon populations are extracted for the $\Delta$, L, and $\Gamma$ valleys of silicon. The results suggest that transient XUV spectroscopy of the Si $L_{2,3}$ edge can be used to resolve different carrier and phonon scattering pathways, but only if the excited state XUV absorption is analyzed in terms of all electronic and structural contributions.

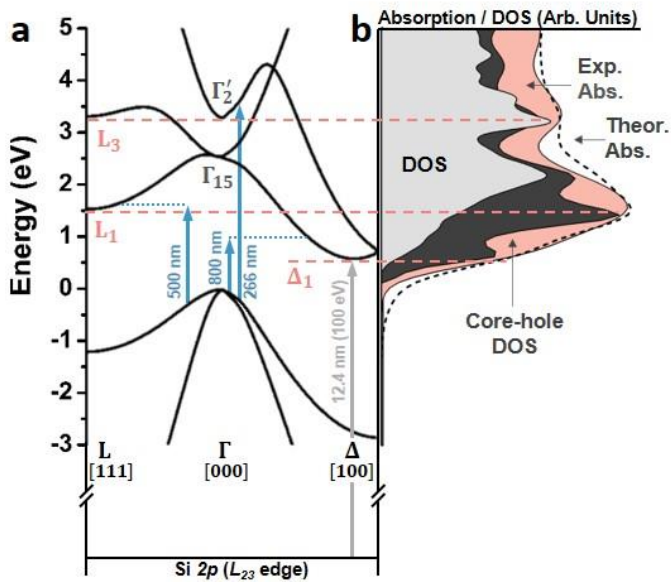

*Figure 1*. *Comparison of the Si band structure to the XUV absorption at the Si $L_{2,3}$ edge. (a) Band structure of silicon along the L-Γ-X path, with L at ½,½,½ and Δ at ~0.8,0,0 in the Brillouin zone. The different photoexcitation energies are marked, including an example of the core level transition. The lengths of the arrows are corrected for the underestimation of the band gap in the*

*DFT calculation. (b) The* s+d *projected density of states (light grey area) and core-hole modified DOS (dark grey area) are compared, as well as the experimental absorption (pink area) and theoretical absorption (black dashed line). The theoretical absorption is predicted by broadening the core-hole modified DOS with a single plasmon pole model. The* s-p *hybridized $L_1$ and $\Delta_1$ critical points are renormalized by the 2p core-hole exciton more so than the d-character $L_3$ critical point. The notation for the critical points is different from the notation for the $L_{2,3}$ x-ray edge.*

## II. RESULTS
### II.1 State-Filling Contributions

The band structure of Si along the L-Γ-Δ path is shown in Figure 1a. The band structure is calculated using DFT (see SI Section 2 for details). The arrows for the photoexcitation energies in Figure 1a are corrected for the underestimated band gap of the DFT calculation to give the proper conduction band energy offsets. As the photoexcitation energy is increased, carriers are promoted to valleys at different k-space positions in the band structure. For 800 nm excitation, an indirect transition is possible to the $\Delta_1$ valley. For 500 nm excitation, an indirect transition is possible to the $L_1$ valley. For 266 nm excitation, a direct transition is possible near the Γ valley in the Γ-X direction[35]. In each case, the photoexcited carriers will eventually thermalize to the conduction band edge ($\Delta_1$) by inter- and intra-valley electron-phonon scattering. The electrons and holes then recombine by Auger and non-radiative recombination.

The Si $L_{2,3}$ edge XUV transition occurs by the promotion of an electron from the Si 2p core level to unoccupied states in the *s*- and *d*-character valence and conduction bands. The Si 2p core level is localized and therefore has a flat dispersion in k-space. Thus, for a given energy, the probability of the Si $L_{2,3}$ transition is the sum over all dipole-allowed transitions across k-space at that energy. After visible light photoexcitation, the change in the Si $L_{2,3}$ edge absorption should represent the change in occupation of the unoccupied states probed. The XUV transition probability will decrease at energies for which electrons have been photoexcited; the XUV transition probability will increase at energies for which photoexcitation leaves behind a photoexcited hole. These changes in the differential absorption are referred to as state-filling effects. Measuring the state-filling in the XUV spectrum should resolve the photoexcited carrier energy and relaxation pathways as a function of time.

The core-hole created in the Si 2p core level, however, obscures the state-filling effects[33]. More specifically, the core-hole potential distorts the final transition density of states from the ground state band structure of Si as shown in Figure 1b. In Figure 1b, the experimental XUV absorption (pink area), which has been deconvoluted for the 0.6 eV $L_{2,3}$ spin orbit splitting, is compared to the partial *s+d* density of states (DOS) calculated using DFT (light grey area)[22]. The core-hole corrected density of states, calculated using the BSE-DFT method, is shown as the dark grey area. The critical points of the ground state band structure are indicated by dashed pink lines. Only when the core-hole effects are included by the BSE do the critical points align with the major peaks of the experimental XUV absorption. Please note, the notation for the critical points is different from the notation for the $L_{2,3}$ x-ray edge.

To experimentally demonstrate the core-hole distortion of the state-filling, the excited state differential absorption spectra are shown in Figure 2a-c, indicating photoexcitation with 800 nm ($\Delta_1$), 500 nm ($L_1$), and 266 nm ($\Gamma_2$') light, respectively. The experimental differential absorption spectra are calculated as the logarithm of the pump-on over the pump-off spectra. Figure 2a-c are plotted as color maps on a logarithmic time axis up to 200 ps, with the blue shade corresponding to an increased absorption and the red shade corresponding to a decreased absorption relative to the ground state. The yellow shade represents no excited state change from the ground state XUV absorption. Further experimental details are given in the Supporting Information Section 1.

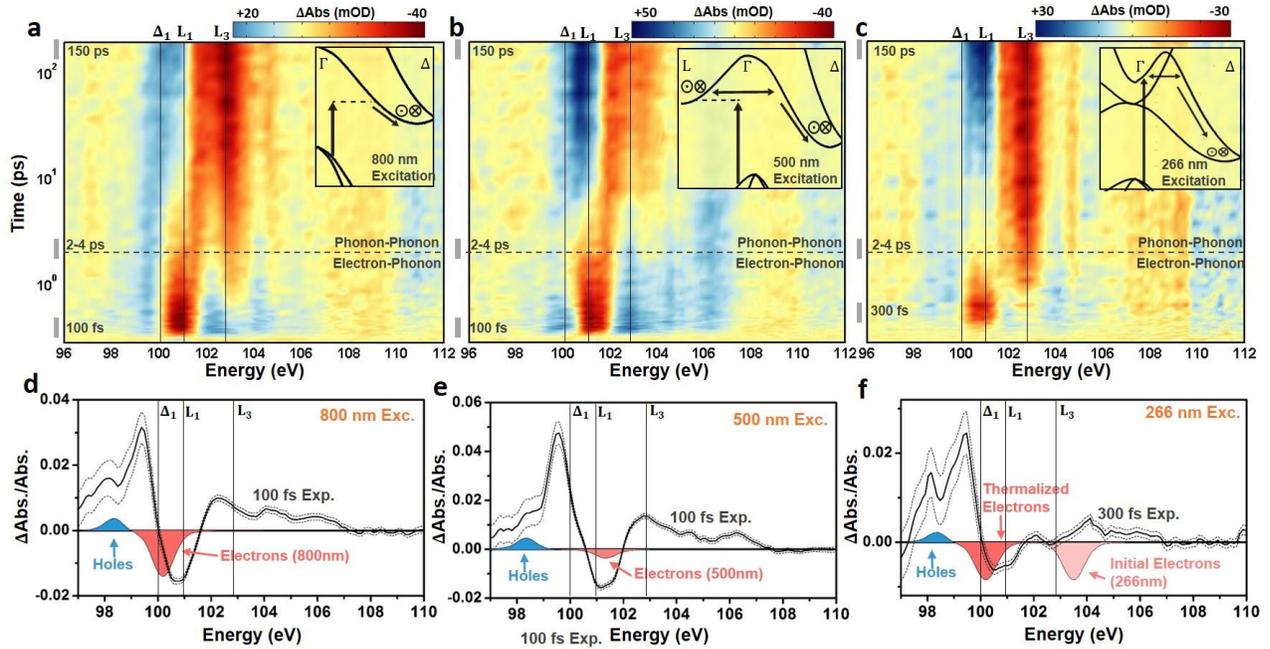

*Figure 2.* Differential absorption spectra of the Si $L_{2,3}$ edge and fit to theory. The differential absorption is shown on a logarithmic time scale from 0 to 200 ps for (a) 800 nm excitation to the $\Delta_1$ valley, (b) 500 nm excitation to the $L_1$ valley, and (c) 266 nm excitation in the $\Gamma$ valley. The cross-over time between predominantly electron-phonon scattering to predominantly phonon-phonon scattering is indicated by the dashed horizontal line. The insets represent some of the possible excitation and scattering pathways for the excited electrons. The in and out of plane arrows indicate where inter-valley scattering between degenerate valleys is possible. (d-f) Comparison of the experimental differential absorption to the predicted state filling effects. The predicted increase in absorption by the photoexcited holes is shown in blue while the decrease in absorption from photoexcited electrons is shown in red. The excitation wavelength is indicated on each panel. The black solid line indicates the experimental differential absorption at 100 fs after photoexcitation, while the black dashed lines indicate the 95% confidence intervals on the experimental data. A 300 fs time slice is used for 266 nm photoexcitation because the electrons are initially excited to the $\Gamma$ valley, which is XUV dipole-forbidden. The energy of the initially excited electrons is represented schematically by the pink area for 266 nm photoexcitation, while the red area represents the modeled state-filling at the $\Delta_1$ point after 300 fs.

Immediately after 800 nm excitation, a decrease in absorption (red) is measured at 100-102 eV, centered between the $\Delta_1$ and $L_1$ critical points, while an increased absorption (blue) is measured at all other energies between 98-105 eV. For 500 nm excitation, the measured decrease in absorption (red) begins at the $L_1$ critical point and spans halfway to the $L_3$ critical point. For 266 nm excitation, the measured decrease in absorption (red) is again centered between the $\Delta_1$ and $L_1$ critical points, but its appearance is delayed in time by several hundred femtoseconds from the pump pulse. For each excitation wavelength, an increase in absorption (blue) is measured at energies below *the Si $L_{2,3}$* edge core-to-valence transition onset energy of 99.8 eV. The transition onset energy should correspond to the conduction band minimum at the $\Delta_1$ point; the valence band is then located at ~98.7 eV for the 1.12 eV band gap of silicon.

Qualitatively, the measured changes in absorption in Figure 2a-c appear to correspond with state-filling. An increase in absorption (blue) from the photoexcited holes is expected around the valence band energy (~98.7 eV) for each photoexcitation wavelength. For 800 nm (Figure 2a), electrons are excited to the $\Delta_1$ valley, so a decrease in absorption (red) is expected at the $\Delta_1$ critical point. In Figure 2b, photoexcitation with 500 nm light to the $L_1$ valley is expected to decrease absorption (red) at the $L_1$ critical point. For 266 nm photoexcitation, electrons are excited to the $\Gamma_2$' point, which is dipole-forbidden for the Si $L_{2,3}$ edge, so an initial decrease in absorption will not be measured[36]. Instead, a delayed decrease in absorption at the $\Delta_1$ critical point is expected on a several hundred femtosecond timescale as $\Gamma$-X inter-valley and then intra-valley electron-phonon scattering thermalize the excited carrier distribution[37,38].

While the measured differential absorptions qualitatively match state-filling, several experimental observables cannot be explained without including the effects of the core-hole perturbation. First, an increased absorption (blue) is measured at energies higher than the $L_3$ critical point for all excitation wavelengths, as well as below the $L_1$ critical point for 500 nm excitation. In a purely state-filling interpretation, the increase in absorption at these energies would have to unphysically correspond to holes being created in already unoccupied states. Second, for 800 nm photoexcitation, the decreased absorption (red) increases in center energy with time toward the $L_1$ critical point between the 100 fs and 2-4 ps lineouts. For 500 nm photoexcitation, the decreased absorption increases in width toward the $\Delta_1$ critical point but does not decrease in center energy on this same timescale. In both cases, photoexcited carriers are thermalizing through electron-phonon scattering during this timescale. Any state-filling contribution should therefore decrease in energy and width, as has been measured with visible and infrared probes. This is in disagreement with the XUV probe measured differential absorption. Finally, on close inspection, the measured decrease in absorption is not energetically aligned with the energy of the photoexcited carriers for each wavelength.

The discrepancy between the expected state-filling effects and the core-hole modified differential XUV absorption is further illustrated in Figure 2d-f. Figure 2d-f compares the core-hole corrected, theoretically predicted differential absorption from state-filling effects (colored areas) to the experimental differential absorption measured immediately following photoexcitation (black line). The predicted blocking of the XUV transitions by photoexcited electrons is shown as a red area, while the predicted opening of new transitions by holes is shown as a blue area. The theoretically predicted state-filling effects are calculated using the experimental excitation density and the core-hole modified DOS as corrected for the dipole-allowed transition amplitudes[22]. For example, the magnitude of the state-filling for the Si $L_{2,3}$ transition is lower for electrons photoexcited to the $L_1$ critical point because the dipole-corrected DOS is larger at this energy. The magnitude of the state-filling is also lower for holes in the majority *p*-characterized valence band. For 266 nm photoexcitation, the thermalized electrons at the $\Delta_1$ point (red) and the initially photoexcited carriers in the $\Gamma_2$' valley (pink) are shown.

The core-hole perturbed state-filling and experimental lineouts in Figure 2d-f confirm that the measured decrease (increase) in absorption from state-filling does not energetically align with the photoexcited electron (hole) distributions expected from the ground state band structure. Additionally, the increase in absorption predicted for the photoexcited holes does not match the width of the measured increase in absorption. The magnitudes of the state-filling contributions also do not match the magnitudes of the total differential absorption signal, nor can the state-filling contributions predict the increased absorption at energies away from the photoexcited critical points. The increased and decreased absorption measured in the core-hole perturbed XUV spectrum are therefore not solely assignable to photoexcited electrons and holes, as is possible in visible light and infrared light experiments.

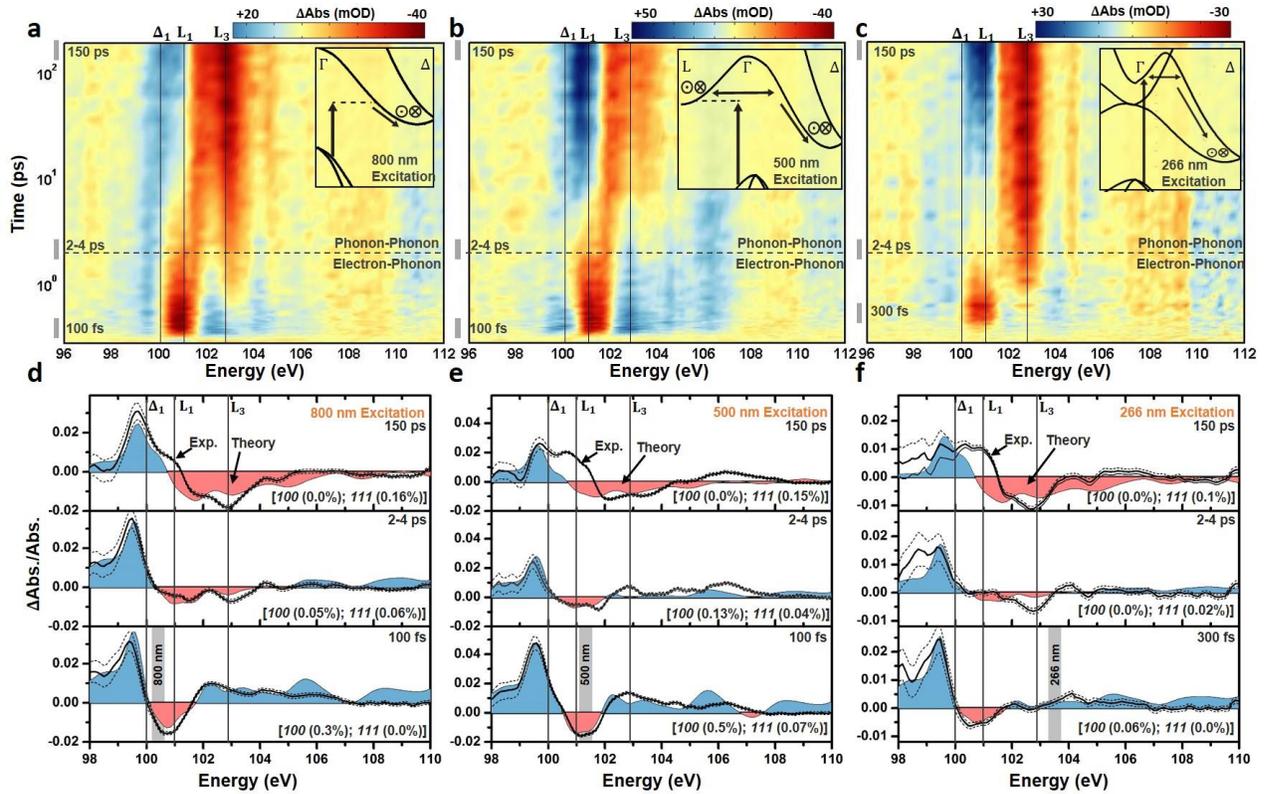

*Figure 3.* Differential absorption spectra of the Si $L_{2,3}$ edge and fit to theory. The differential absorption is shown on a logarithmic time scale from 0 to 200 ps for (a) 800 nm excitation to the $\Delta_1$ valley, (b) 500 nm excitation to the $L_1$ valley, and (c) 266 nm excitation in the $\Gamma$ valley. The cross-over time between predominantly electron-phonon scattering to predominantly phonon-phonon scattering is indicated by the dashed horizontal line. The log scale of time is offset by 100 fs for visualization. The insets represent some of the possible excitation and scattering pathways for the excited electrons. The in and out of plane arrows indicate where inter-valley scattering between degenerate valleys is possible. Time slices (times indicated by gray bars to the left of panels a-c) from the different scattering periods is shown in (d)-(f). The theoretical predictions from the single plasmon pole and BSE-DFT calculation, described in the SI Section 2, are shown as the blue and red colored areas. The best fit percentage expansion of the relevant lattice vectors is indicated on the bottom right of each panel. In (d)-(f) the $\Delta Abs./Abs.$ scale is used to allow direct comparison of experiment (solid line) to theory (blue and red areas) without scaling of the results. The dashed lines show the 95% confidence intervals on the experimental data. The lineouts in Figure 3d-f are averaged over the four nearest time-points.

## II.2 Other Electronic and Lattice Contributions

After photoexcitation, a range of electronic and structural changes occur to the Si $L_{2,3}$ edge on timescales from femtoseconds to hundreds of picoseconds[33]. The electronic changes can be summarized as renormalization of the band gap, changes in excited state broadening, and state-filling effects. The anharmonicity of any phonons excited by electron-phonon and phonon-phonon scattering, as well as the photoexcited screening of the Si-Si bonds, results in structural changes. To approximate the possible core-hole modified electronic and structural effects, the model of Reference 33 is employed as outlined in the Supporting Information Section 2. As shown in Figure 3 and discussed below, inclusion of the other electronic and structural effects in addition to the state-filling effects leads to a better theoretical description of the XUV transient absorption spectra.

Briefly, in Figure 3d-f, the experimental excitation density is used to estimate the screening of the XUV transition, the excited state broadening, and the state-filling effects. These changes are included in the BSE-DFT calculation of the ground state XUV absorption by energetically shifting the conduction band edge, changing the carrier density in the plasmon frequency of the Drude-Lindhard broadening model, and including the change in occupation of the core-hole modified DOS. The excited-state carrier distribution is modeled as a Gaussian with width 0.3 eV, corresponding to the experimental broadening, and with an initial energy determined by the photoexcitation energy. Structural effects are included by best-fitting a range of equilibrium lattice expansions and strains to the residual experimental differential absorption not predicted by the electronic effects. The spectra are best fit by structural contributions from a [100] lattice deformation and an isotropic lattice expansion (denoted as *111* for convenience in figures). Other combinations of biaxial and triaxial strain could not fit the spectrum.

**Table 1** *Non-linear Fit Carrier Densities and Lattice Expansions. Fit quantities are for an average of the four times around the time indicated.* 100 *expansion denotes an expansion along the [100] direction while* 111 *expansion denotes an isotropic expansion along each axis.*

|  |  | ~100 fs (300 fs[a]) | ~2-4 ps | ~150 ps |
|---|---|---|---|---|
| **800 nm (Δ)** | **Carrier Density (x$10^{20}$/cm$^3$)** | 1.5 ± 0.1 | 1.5 ± 0.1 | 0.6 ± 0.1 |
|  | *100* **Expansion (%)** | 0.3 ± 0.05 | 0.05 ± 0.03 | 0.00 ± 0.08 |
|  | *111* **Expansion (%)** | 0.0 ± 0.01 | 0.06 ± 0.01 | 0.16 ± 0.02 |
| **500 nm (L)** | **Carrier Density (x$10^{20}$/cm$^3$)** | 1.7 ± 0.1 | 1.0 ± 0.1 | 0.8 ± 0.2 |
|  | *100* **Expansion (%)** | 0.5 ± 0.05 | 0.13 ± 0.04 | 0.0 ± 0.1 |
|  | *111* **Expansion (%)** | 0.07 ± 0.01 | 0.04 ± 0.01 | 0.15 ± 0.03 |
| **266 nm (Γ)** | **Carrier Density (x$10^{20}$/cm$^3$)** | 0.9 ± 0.1 | 0.5 ± 0.1 | 0.2 ± 0.1 |
|  | *100* **Expansion (%)** | 0.06 ± 0.02 | 0.00 ± 0.03 | 0.0 ± 0.1 |
|  | *111* **Expansion (%)** | 0.00 ± 0.01 | 0.02 ± 0.01 | 0.10 ± 0.02 |

[a]*300 fs for 266 nm excitation*

The predicted differential absorption is compared to the measured differential absorption at three key timescales in Figure 3d-f. The fit parameters for the three timescales and each excitation wavelength are summarized in Table 1. Immediately following photoexcitation (bottom panels in Figure 3d-f, 100 fs or 300 fs) the differential absorption is best fit for 800 nm, 500 nm, and 266 nm excitation by a carrier density of 1.5±0.1 x$10^{20}$, 1.7±0.1 x$10^{20}$, and 0.9±0.1 x$10^{20}$ carriers/cm$^3$, respectively. For 266 nm excitation, the carrier distribution is taken as thermalized to the $\Delta_1$ point by 300 fs. A [100] lattice expansion of 0.3±0.05%, 0.5±0.05%, and 0.06±0.02% is required for 800 nm, 500 nm, and 266 nm excitation, respectively. For 500 nm excitation, an isotropic expansion of 0.07±0.01% is also fit. An isotropic expansion is not fit for 800 nm or 266 nm excitation within a ±0.01% error. The major discrepancies between the fit and measured differential absorption occur at the critical points that are not accurately captured in the ground state absorption model.

At 2-4 ps after photoexcitation (shown in the middle panel in Figure 3d-f), the experimental differential absorption is best fit for 800 nm, 500 nm, and 266 nm excitation by a carrier density of 1.5±0.1 x$10^{20}$, 1.0±0.1 x$10^{20}$, and 0.5±0.1 x$10^{20}$ carriers/cm$^3$, respectively. The electrons are taken as thermalized to the $\Delta_1$ point for each excitation wavelength. The [100] lattice deformation fitted amplitudes decrease from the 100 fs value for all excitation wavelengths. The fit values decrease from 0.3±0.05% to 0.05±0.03% for 800 nm photoexcitation, 0.5±0.05% to 0.13±0.04% for 500 nm photoexcitation, and 0.06±0.02% to 0.00±0.03% for 266 nm photoexcitation. A lattice expansion of 0.06±0.01% and 0.02±0.01% is fit for 800 nm and 266 nm excitation, respectively. For 500 nm excitation, the lattice expansion coefficient decreases from 0.07±0.01% to 0.04±0.01%. The change in the isotropic lattice expansion coefficient could correspond to the anharmonicity of acoustic phonons excited by the onset of intra-valley electron-phonon scattering as well as phonon-phonon decay processes[39–41]. This assignment is supported by the decay of the [100] lattice deformation component, which is attributed to either the excited optical phonon population or the anisotropy of the carrier distribution. Given this expansion is faster than the acoustic velocity, it is most likely a combination of volume-preserving deformations or disordering[42].

At 150 ps after photoexcitation (top panels in Figure 3d-f), the experimental differential absorption for each photoexcitation wavelength converges to a similar spectrum. Below 101-102 eV there is an increase in absorption (blue); above 101-102 eV there is a decrease in absorption (red). The lineouts at 150 ps can be fit to a carrier density of 0.6±0.1 x$10^{20}$, 0.8±0.2 x$10^{20}$, and 0.2±0.1 x$10^{20}$ carriers/cm$^3$ for 800 nm, 500 nm, and 266 nm excitation, respectively. The electrons are again assumed to be thermalized to the $\Delta_1$ point. A [100] lattice deformation is not required to fit the experimental data within the ±0.1% error. An isotropic lattice expansion of 0.16±0.02%, 0.15±0.03%, and 0.10±0.02% is required for 800 nm, 500 nm, and 266 nm excitation, respectively. The decreased carrier densities are indicative of Auger recombination[43]. The increased lattice expansions are indicative of the thermalizing phonon bath and the lattice becoming heated[31,39–41]. The crossover energies between the areas of increased and decreased absorption vary slightly between the excitation energies in the top panels of Figure 3d-f, as does the amplitude of the increased absorption at the $L_1$ critical point. These changes are not captured by the model, suggesting that the final lattice distortion cannot be completely described by the three lattice distortions that were tried, namely [100], [110], and [111] lattice distortions.

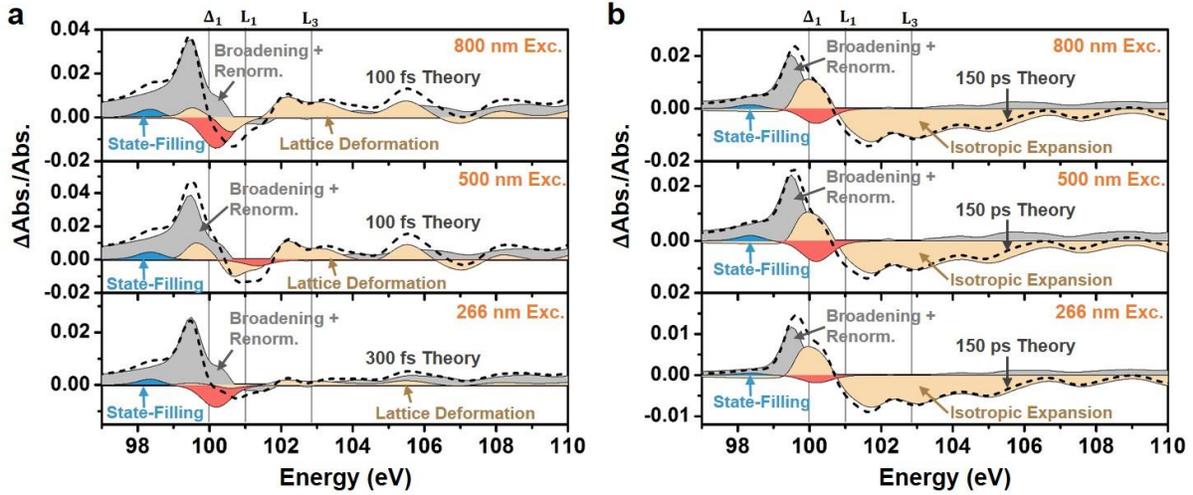

***Figure 4.*** *Theoretical decomposition of the differential absorption at the Si $L_{2,3}$ edge. The relative contributions of state-filling (blue area for holes, red area for electrons), broadening and renormalization (grey area), and the lattice deformation and expansion (orange area) in the theoretically predicted differential absorption are shown at (a) 100 fs and (b) 150 ps after excitation. For 266 nm excitation, a 300 fs plot is shown in (a) because the XUV transition to the initially occupied valley is dipole forbidden. The black dashed line indicates the sum of the predicted components and is the same as the best-fit in Figure 3d-f. The excitation wavelength is indicated on each plot. At 100 fs after excitation, the increased and decreased absorption regions are a mix of electronic and structural components and cannot be assigned purely to state-filling from photoexcited carriers. At 150 ps after excitation, the differential absorption is dominated by the heating of the lattice.*

**II.3 Interpretation of Spectra including Core-Hole Effects**

Using the full model of Section II.2, the best-fits of the experimental differential absorptions can be decomposed into the amplitudes of the different core-hole modified electronic and structural effects (Figure 4). In this manner, the spectral features of the experimental differential absorption can be approximately assigned. Immediately following photoexcitation (Figure 4a), the increase in absorption below the transition edge is primarily from broadening and renormalization (grey area), matching Reference 33. The state-filling from holes (blue area) and structural changes (orange area) play more minor roles. The excited-state renormalization and broadening create a large below-edge differential absorption because the red-shift of the XUV spectrum increases absorption at energies where absorption was previously not possible. For energies between the $\Delta_1$ and $L_1$ critical points, the measured decrease in absorption is caused by an equal mixture of all four effects. The state-filling from electrons (red area) cancels with the broadening and renormalization (grey area) near the $\Delta_1$ point. This explains why the state-filling component for electrons does not energetically align with the measured decrease in absorption unless all core-hole perturbations of the final state are approximated. At energies above the $L_3$ critical point, the differential absorption is a positive amplitude mixture of broadening, renormalization, and structural components.

As the photoexcitation energy is increased, the center energy of the state-filling component increases immediately following photoexcitation (top to bottom panel in Figure 4a). In the model used, the broadening and renormalization from the core-hole are unchanged as the excitation energy changes because they only depend on the carrier density. Comparing the 800 nm and 500 nm panels in Figure 4a, the change in the differential absorption between the $\Delta_1$ and the $L_1$ critical points is because of the change in the electron state-filling energy. The increased center energy of the photoexcited electrons shifts the decreased absorption feature to the $L_1$ critical point. However, this shift in center energy also decreases the cancellation at the $\Delta_1$ point, effectively increasing the absorption. The structural components also contribute to the measured differences in absorption around these critical points. Therefore, when core-hole effects are included, the measured differences between 800 nm and 500 nm photoexcitation are related to the difference in state filling, but the decreased absorption cannot be assigned only to the excited electron energy.

By 150 ps after photoexcitation (Figure 4b), all carriers are thermalized to the $\Delta_1$ point and Auger recombination has decreased the photoexcited carrier density[43]. This is evidenced by a reduced magnitude of the positive differential absorption at energies below the $L_1$ critical point. At energies above the $L_1$ critical point, the predicted differential absorption is dominated by the isotropic expansion from the heated lattice. At the $\Delta_1$ critical point, the lattice expansion and state filling components have similar amplitude but opposite sign. This cancellation is why the decreased absorption signature in Figure 3a-c disappears on a timescale faster than expected for recombination.

In summary, Figure 4 shows that the core-hole must be considered to accurately map spectral changes in the XUV spectrum to each spectral change expected from visible and infrared probe experiments. This includes correcting the final transition density of states, the renormalization with carrier density, the XUV broadening, and the spectral appearance of structural modes in the XUV spectrum. Within the assumptions of the model used, the features of the >100 ps differential absorption spectrum is attributed primarily to the heated lattice, whereas the initial transient XUV signal is a mixture of electronic and structural effects. Immediately after photoexcitation, the primary differences between the excitation wavelengths occur because of the energy of the photoexcited electrons. Between 100 fs and 2-4 ps, thermalization of the photoexcited electrons, and the lattice deformations that result, create the differences in the measured differential absorption. After 150 ps, the electrons are fully thermalized to the conduction band edge, and the measured differential absorption spectra converge because the primary contribution to the XUV spectrum is the isotropic expansion of the heated lattice. For each photoexcitation wavelength, the differential absorption above the $L_3$ critical point primarily corresponds to structural modes, while the differential absorption at energies below the $L_3$ critical point is a mixture of electronic and structural components. These findings are consistent with the analysis of the static x-ray absorption in terms of near-edge and far-edge features in synchrotron studies.

## III. DISCUSSION

The core-hole effects mean that the features of the experimental differential absorption are not solely assigned to state-filling from photoexcited electrons and holes that results from their photoexcitation (Figure 4). The core-hole perturbation also means that structural changes are exaggerated in the differential spectrum because of the core-hole's localized wave function. The model used to predict core-hole effects in this paper is an approximate approach to treating the core-hole perturbation. To test the accuracy of this model[33], the magnitude of each electronic and structural component was fit as a function of time. This allows the extracted dynamics to be compared to those known from visible light and infrared experiments. In this fit, the fit parameters are the carrier density and electron center energy, as well as the structural components which give insight into the inter-valley, intra-valley, and phonon-phonon decay processes.

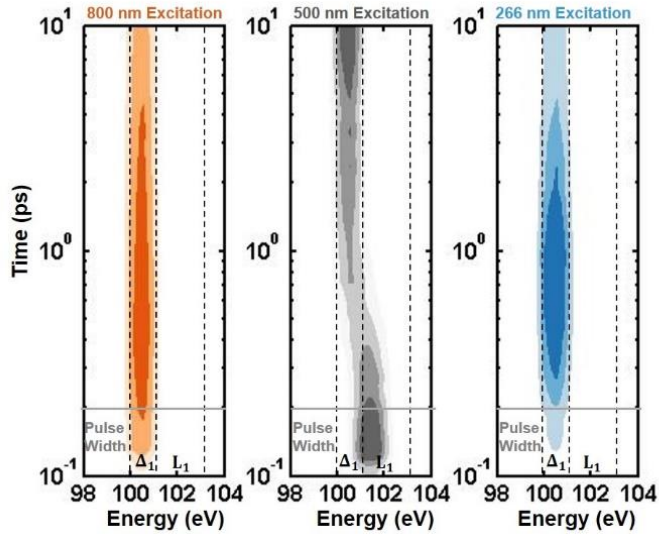

*Figure 5. Extracted electron distribution from the Si $L_{2,3}$ edge differential absorption using the BSE-DFT model. The color scale in each panel corresponds to the qualitatively higher (darker) and lower (lighter) extracted carrier density. For 800 nm excitation, carriers are excited into the $\Delta_1$ valley within the pulse width and then thermalize and recombine. For 500 nm excitation, carriers are excited into the $L_1$ valley and then thermalize to the $\Delta_1$ valley. The amplitude has a minimum before 10 ps because of the dipole selection rules near the $\Gamma$ valley. For 266 nm excitation carriers are excited into the $\Gamma$ valley, which cannot be probed by the Si 2p core level because of dipole selection rules, and then transfer to and thermalize in the $\Delta_1$ valley within 300 fs. The log timescale is offset by 100 fs for visualization and extends to 10 ps. The experimental and excited state broadening limits the intra-valley energetic thermalization that can be observed. The color scales are normalized.*

The extracted electron state-filling from fitting the experimental differential absorption to the model of Reference 33 is shown in Figure 5. For 800 nm excitation, the experimental and excited state broadening prevents the thermalization of the electrons from being observed at a resolution better than ~0.5 eV. The amplitude of the state-filling otherwise follows the expected intra-valley thermalization and Auger decay timescales[43]. For 500 nm excitation, the photoexcited electrons initially occupy the $L_1$ critical point, then relax to the $\Delta_1$ valley on the timescale of the electron-phonon thermalization[37,38]. The middle panel of Figure 5 shows that the model extracts this trend. A notable feature is the difference in amplitude as carriers fill the smaller joint DOS for the Si 2p transition to the $\Delta_1$ versus the $L_1$ critical point, as well as when they first scatter in to the $\Delta$ valley near the $\Gamma$ point around 10 ps. For 266 nm excitation, a delayed rise is observed in the probed $\Delta_1$ critical point, consistent with $\Gamma$-X direction inter-valley scattering followed by intra-valley scattering on a hundreds-of-femtosecond timescale. On a picosecond timescale, the amplitude of the state-filling decreases, consistent with Auger recombination and thermalization time-scales[43]. For 266 nm excitation, state-filling is not fit in the L valleys, consistent with a stronger $\Gamma$-X direction than $\Gamma$-L direction inter-valley scattering[35,37,38].

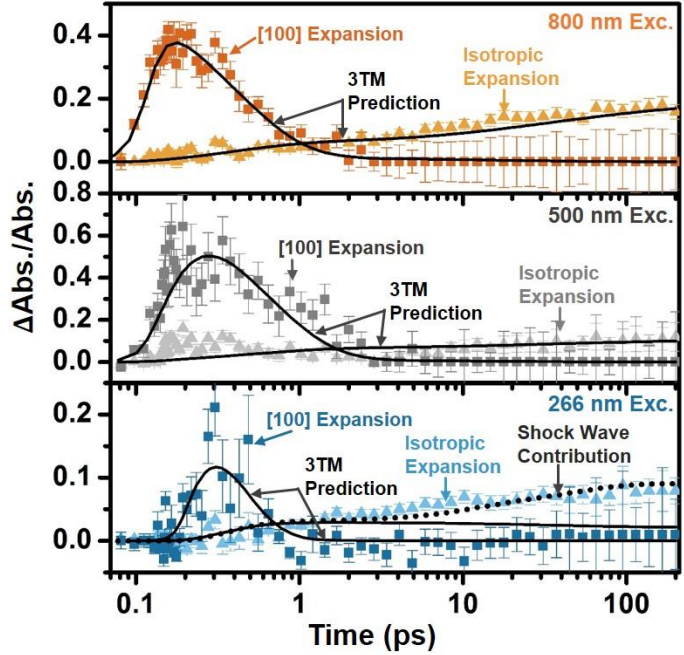

*Figure 6.* Best fit lattice expansion coefficients for 800 nm, 500 nm, and 266 nm excitation. The squares correspond to the [100] lattice deformation component. The triangles correspond to the isotropic lattice expansion component. The black solid line corresponds to a fitting of the extracted structural components with a three-temperature model (3TM). The trade-off between the [100] expansion and the isotropic expansion with increasing time is indicative of the switch from carrier thermalization to lattice heating. The three-temperature model describes this trade-off through fitting the average electron-optical phonon, electron-acoustic phonon, and phonon-phonon lifetimes as reported in Table 2. The dotted line shows the predicted shock-wave contribution from the spatially-variant carrier profile. The excitation wavelengths are otherwise indicated on the plot, and the error bars correspond to the standard error from a robust non-linear fit weighted by the experimental uncertainty. The experimental noise level is reflected in the scatter of the data points. The logarithmically scaled time axis is offset by 100 fs for visualization.

**Table 2** *Non-linear Fit Carrier and Phonon Scattering Times. The coefficient values are determined by fitting the three-temperature model (3TM) to the structural components extracted from the experimental differential absorption using the BSE-DFT model.*

|  |  | Electron-Optical Phonon | Electron-Acoustic Phonon | Optical Phonon–Acoustic Phonon |
|---|---|---|---|---|
| 800 nm ($\Delta$) | $\tau$ (fs) | 30 ± 10 | 400 ± 10 | 470 ± 20 |
| 500 nm (L) | $\tau$ (fs) | 90 ± 30 | 300 ± 100 | 600 ± 100 |
| 266 nm ($\Gamma$) | $\tau$ (fs) | 110 ± 20 | 570 ± 20 | 180 ± 20 |

The structural components determined by the best-fit of the model to the experimental data are shown as a function of time in Figure 6. The squares represent the [100] lattice deformation component, whereas the triangles represent the isotropic lattice expansion component. The extracted amplitudes are fit to a three-temperature model (solid line) to test if the kinetics of the structural components match the known scattering processes within each photoexcited valley. The 3TM, described in detail in the Supporting Information Section 3, can represent the average inter-valley electron-phonon ($\tau_{e-o}$), intra-valley electron-phonon ($\tau_{e-a}$), and phonon-phonon ($\tau_{o-a}$) scattering processes previously reported from visible light and infrared experiments. For silicon, the inter-valley electron-phonon scattering mostly involves high energy optical phonon modes (e-o), while the intra-valley scattering mostly involves acoustic phonon modes (e-a)[44–46]. The 3TM is chosen because it allows for the kinetics of the extracted structural components to be quantified while also accurately modeling the absorption depth, carrier diffusion, and shock wave contributions for the range of excitation wavelengths used in the experiments. The fit coefficients are summarized in Table 2.

For 800 nm excitation, the structural coefficient amplitudes are fit to time constants of 30±10 fs for $\tau_{e-o}$, 400±10 fs for $\tau_{e-a}$, and 470±20 fs for $\tau_{o-a}$. These time constants match previous reports of silicon's inter-valley scattering time of 20-60 fs[47,48], electron-acoustic phonon scattering time of 500 fs[49], and the 400 fs screened optical phonon lifetime at $10^{20}$ carriers/cm$^3$ excitation[50]. For 500 nm excitation, the scattering times are 90±30 fs for $\tau_{e-o}$, 300±100 fs for $\tau_{e-a}$, and 600±100 fs for $\tau_{o-a}$. The 3TM does not include the possibility of two different early timescale structural contributions, resulting in the 100 fs fit error for the intra-valley and phonon-phonon scattering times. The inter-valley scattering time $\tau_{e-o}$ correlates well with a mixture of L-L' inter-valley scattering on a 20-100 fs timescale and the slightly slower 100-200 fs timescale of L-X inter-valley scattering[37,38]. Within the fit error, a similar intra-valley scattering time $\tau_{e-a}$ and screened phonon-phonon scattering time $\tau_{o-a}$ are fit at 500 nm as at 800 nm excitation. The inter-valley and intra-valley scattering times match the thermalization of the electron energy in Figure 6.

For 266 nm excitation, the fit time constants are 110±20 fs for $\tau_{e-o}$, 570±20 fs for $\tau_{e-a}$, and 180±20 for $\tau_{o-a}$. The slower inter-valley scattering time compared to 500 nm or 800 nm excitation is consistent with the slower inter-valley electron-phonon scattering expected from the smaller $\Gamma_2'$-X joint DOS[37,38]. The longer intra-valley scattering time is also consistent with the increased excess energy that must be dissipated after $\Gamma$-X inter-valley scattering. However, several discrepancies exist in the 3TM fit. First, the 3TM cannot replicate the long timescale kinetics of the isotropic lattice expansion. Second, the phonon-phonon decay time is reduced by several hundred femtoseconds compared to 800 nm or 500 nm excitation. These trends must be corrected for the 5-10 nm absorption depth of 266 nm light in silicon as follows[51].

The small absorption depth compared to the sample thickness creates a photoexcited carrier distribution gradient. The spatial carrier gradient excites an acoustic shock wave, which can be modeled using the elasticity equation [40,52–56]

$$\rho v^2 \frac{\partial^2 u}{\partial z^2} = \rho \frac{\partial^2 u}{\partial t^2} + F(z,t), \qquad (5)$$

where $\rho = 2.328 \ g/cm^3$ is the density of Si, $u$ is the displacement, $v = 8430 \ m/s$ is the longitudinal sound velocity in silicon[57], and $F(z,t) = \frac{\partial \sigma(z,t)}{\partial z}$ is the source term with the time dependent stress calculated using the spatial and temporal results of the 3TM. An acoustic shock wave contribution with this velocity will become noticeable 1 ps after excitation (wave moves ~10 nm) and reach a spatially-averaged maximum in 20-30 ps for a 200 nm membrane. As shown by the dotted line in the bottom panel of Figure 6, the acoustic shock wave increases the agreement between the 3TM prediction and the experimental data at time-scales longer than 10 ps. It should be noted that the transmission geometry makes the probe spectrum, and therefore Figure 6, more representative of the post-shock-wave expansion than the wave-front shock region, which is composed of positive and negative strain regions of similar magnitude[40].

The spatial carrier gradient also means that the peak carrier density at the surface is higher than the spatially-averaged value calculated from Equation (1). The peak carrier density will approach the spatially-averaged carrier density as carriers diffuse away from the surface, as captured by the 3TM in Figure 6 and shown in Figure 7. Immediately following photoexcitation, the peak carrier density is up to ten times higher than the spatially averaged value present for 500 nm or 800 nm excitation. Extrapolating the data of References 50 and 58, this increase in peak carrier density would results in a screened phonon-phonon decay time of 200-300 fs. The spatial carrier gradient may therefore explain the smaller optical phonon decay time fit for 266 nm excitation. Of course, as is the case at the other excitation wavelengths, the 3TM is a simplification of the complex inter- and intra-valley thermalization following 266 nm excitation. Figure 6 and 7 should therefore only be considered as an approximate indicator of an acoustic shock wave and carrier-gradient effects.

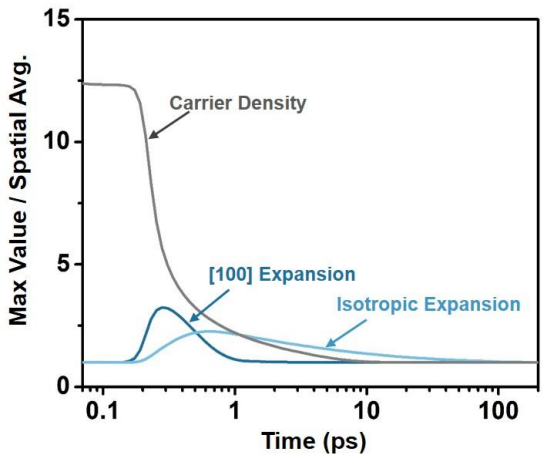

*Figure 7. Effect of the spatially-variant carrier distribution excited by 266 nm light. Solid lines indicate the ratio of the maximum spatially-variant quantities to the spatially-averaged values for the carrier density, [100] lattice expansion, and isotropic expansion calculated using the three temperature model (3TM). The log scale of time is offset by 100 fs for visualization.*

## IV. CONCLUSION

In conclusion, the transient XUV signal of the silicon 2p $L_{2,3}$ edge is measured following excitation to the Δ, L, and Γ valleys of silicon. The experimental transient XUV differential absorption could not be interpreted solely in terms of the state-filling effects from the photoexcited electrons and holes expected from visible and infrared light probe measurements. Instead, the core-hole perturbation to the electronic and structural contributions must be modelled in the differential absorption. To approximate these many body effects, the data is analyzed in terms of a ground state BSE-DFT calculation of the silicon XUV absorption. The BSE-DFT model is used to assign the measured differential absorption features. Immediately after photoexcitation, a mixture of core-hole modified effects from state-filling, excited state broadening and renormalization, and lattice deformation is predicted to lead to the measured spectrum. On the hundreds-of-picoseconds timescale, the measured differential absorption is mainly the result of the heated lattice, the effects of which are prevalent in the differential absorption because of the localized core-hole wave function. The validity of the BSE-DFT model is tested by fitting the electron energy as a function of time from the experimental data. When corrected for the possible dipole transitions, the carrier thermalization from the L and Γ valleys to the Δ valley is quantified, but the energetic resolution is limited by the core-hole broadening. The fit kinetics of the structural modes also match the electron-phonon and phonon-phonon scattering times previously reported for the Δ, L, and Γ valleys of silicon. These results show that transient XUV spectroscopy of the Si $L_{2,3}$ edge cannot be interpreted without modeling the core-hole perturbation to some degree. The model of Reference 33, although approximate, was found to extract the photoexcited dynamics with timescales that match previously reported carrier thermalization pathways.

**SUPPORTING INFORMATION**

Further information on the experimental and theoretical methods are included in the Supporting Information.

**ACKNOWLEDGEMENTS**

This work is supported by the "Physical Chemistry of Inorganic Nanostructures Program, KC3103, Office of Basic Energy Sciences of the United States Department of Energy under Contract No. DE-AC02-05CH11231. SKC acknowledges support by the Department of Energy, Office of Energy Efficiency and Renewable Energy (EERE) Postdoctoral Research Award under the EERE Solar Energy Technologies Office. H.-T. C. acknowledges support by the Air Force Office of Scientific Research (AFOSR) (FA9550-15-1-0037). M.Z. acknowledges support from the Humboldt Foundation.

**TOC Graphic**

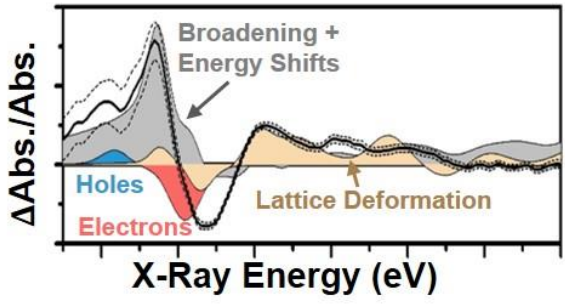